\documentclass[conference]{IEEEtran}
\IEEEoverridecommandlockouts
\usepackage{amsfonts}
\usepackage{amsmath}
\usepackage{balance}
\usepackage{hyperref}

\usepackage{ifpdf}
\ifCLASSINFOpdf
   \usepackage[pdftex]{graphicx}
   \graphicspath{{./pdf/}}
   \DeclareGraphicsExtensions{{.pdf}}
\else
   \usepackage[dvips]{graphicx}
   \graphicspath{{./eps/}}
   \DeclareGraphicsExtensions{{.eps}}
\fi

\begin{document}

\title{A New Analysis of the DS-CDMA \\ Cellular Uplink Under Spatial Constraints}
\author{
\IEEEauthorblockN{Don Torrieri,\IEEEauthorrefmark{2}
Matthew C. Valenti,\IEEEauthorrefmark{1}
and Salvatore Talarico\IEEEauthorrefmark{1} }
\IEEEauthorrefmark{2}U.S. Army Research Laboratory, Adelphi, MD, USA. \\
\IEEEauthorblockA{\IEEEauthorrefmark{1}West Virginia University, Morgantown, WV, USA.
}
\vspace{-0.35cm}
\thanks{M.C. Valenti and S. Talarico were sponsored by the National Science Foundation under Award No. CNS-0750821.}
}
\date{}
\maketitle

\begin{abstract}
A new analysis is presented for the direct-sequence code-division multiple
access (DS-CDMA) cellular uplink. For a given network topology, closed-form
expressions are found for the outage probability and rate of each uplink in
the presence of path-dependent Nakagami fading and log-normal shadowing. The
topology may be arbitrary or modeled by a random spatial distribution for a
fixed number of base stations and mobiles placed over a finite area with the
separations among them constrained to exceed a minimum distance. The
analysis is more detailed and accurate than existing ones and
facilitates the resolution of network design issues, including the influence
of the minimum base-station separation, the role of the spreading factor, and
the impact of various power-control and rate-control policies. It is shown
that once power control is established, the rate can be allocated according
to a fixed-rate or variable-rate policy with the objective of either meeting
an outage constraint or maximizing throughput. An advantage of the variable-rate
policy is that it allows an outage constraint to be enforced on every
uplink, whereas the fixed-rate policy can only meet an average outage constraint.
\end{abstract}

\vspace{-0.1cm}

\thispagestyle{empty}

\section{Introduction}

The classical analysis of the uplinks $\left( \text{e.g., \cite{gil}},\text{
\cite{vit}, \cite{zor}}\right) $ in a cellular network entails a number of
questionable assumptions. Among the principal ones are the existence of a
lattice or regular grid of base stations and the modeling of intercell
interference at a base station as a fixed fraction of the total
interference. Although conceptually simple and locally tractable, the grid
assumption is a poor model for actual base-station deployments, which cannot
assume a regular grid structure due to a variety of regulatory and physical
constraints. The intercell-interference assumption is untenable because the
fractional proportion of intercell interference varies substantially with
the mobile and base-station locations, the shadowing, and the fading.

More recent analyses of cellular networks (e.g., \cite{and}, \cite{Dhillon:2012})
locate the mobiles and/or base stations according to a
two-dimensional Poisson point process (PPP) over a network that extends
infinitely on the Euclidian plane, thereby allowing the use of analytical
tools from stochastic geometry \cite{stoy}, \cite{bacc}. Although the two
principal limitations of the classical approach are eliminated, the PPP
approach is still unrealistic because it cannot model networks of finite
area and does not permit a minimum separation between base stations or between
mobiles, although both minimum separations are characteristic of actual macro-cellular deployments.

The analysis in this paper is driven by a new closed-form expression \cite%
{tor2} for the\ \emph{conditional} outage probability of a communication
link, where the conditioning is with respect to an arbitrary realization of
the network topology. The approach involves drawing realizations of the
network according to the desired spatial and shadowing models, and then
computing the outage probability. Because the outage probability is averaged
over the fading, it can be found in closed form with no need to simulate the
corresponding channels. A Nakagami-m fading model is assumed, which models a
wide class of channels, and the fading parameters do not need to be
identical for all communication links. This flexibility allows the modeling
of distance-dependent fading, where mobiles close to the base station have a
dominant line-of-sight (LOS) path, while the more distant mobiles are
non-LOS. The analysis can characterize more than just the average
performance of a typical uplink.
The outage probability of each uplink can be constrained, and the statistics
of the rate provided to each mobile for its uplink can be determined under
various power-control and rate-control policies.

Although our analysis is applicable for arbitrary topologies, we focus on a
constrained random spatial approach to model the direct-sequence
code-division multiple access (DS-CDMA) uplink. The spatial model places a
fixed number of base stations within a region of finite extent. The model
enforces a minimum separation among the base stations for each \emph{network
realization}, which comprises a base-station placement, a mobile placement,
and a shadowing realization. The model for both mobile and base-station
placement is a binomial point process (BPP) with repulsion, which we call a
\emph{uniform clustering} model \cite{tor2}.

A DS-CDMA uplink differs from a downlink, which has been presented in a
separate paper \cite{val}, in at least four significant ways. First, the
sources of interference are many mobiles for the uplink whereas the sources
are a few base stations for the downlink. Second, sectorization is a
critical factor in uplink performance whereas it is of
minor importance in downlink performance. Third, the uplink signals arriving
at a base station are asynchronous and, hence, non-orthogonal.
Fourth, base stations are equipped with better transmit high-power amplifiers and receive low-noise amplifiers than the mobiles. The net result is that the operational signal-to-noise ratio is typically 5 - 10 dB lower for the uplink than for the downlink.

\section{Network Model}


\label{Section:SystemModel} Cells may be divided into \textit{sectors} by
using several directional sector antennas or arrays, each covering disjoint
angular sectors, at the base stations. Only mobiles in the directions
covered by a sector antenna can cause multiple-access interference on the
\textit{uplink} from a mobile to its associated sector antenna. Thus, the
number of interfering signals on an uplink is reduced approximately by a
factor equal to the number of sectors. Practical sector antennas have
patterns with sidelobes that extend into adjacent sectors, but the
performance degradation due to overlapping sectors is significant only for a
small percentage of mobile locations. Three ideal sector antennas and
sectors per base station, each covering $2\pi /3$ radians, are assumed in
the subsequent analysis.\ The mobile antennas are assumed to be
omnidirectional.

The network comprises $C$ base stations and cells, $3C$ sectors $%
\{S_{1},...,S_{3C}\},$
and $M$
mobiles $\{X_{1},...,X_{M}\}$. The base stations and mobiles are confined to
a finite area, which is assumed to be a circle of radius $r_{\mathsf{net}}$
and area $\pi r_{\mathsf{net}}^{2}$.
The sector boundary angles are the same for all base stations. The
variable $S_{j}$ represents both the $j^{th}$ sector antenna and its
location, and $X_{i}$ represents the $i^{th}$ mobile and its
location. An \emph{exclusion zone} of radius $r_{\mathsf{bs}}$ surrounds
each base station,
and no other base stations or mobiles are allowed within this zone. Similarly,
an exclusion zone of radius $r_{\mathsf{m}}$ surrounds each mobile where no
other mobiles are allowed.

\begin{figure}[t]
\centering

{\begin{tabular}{cc}
\hspace{-0.5cm}
\includegraphics[width=4.45cm]{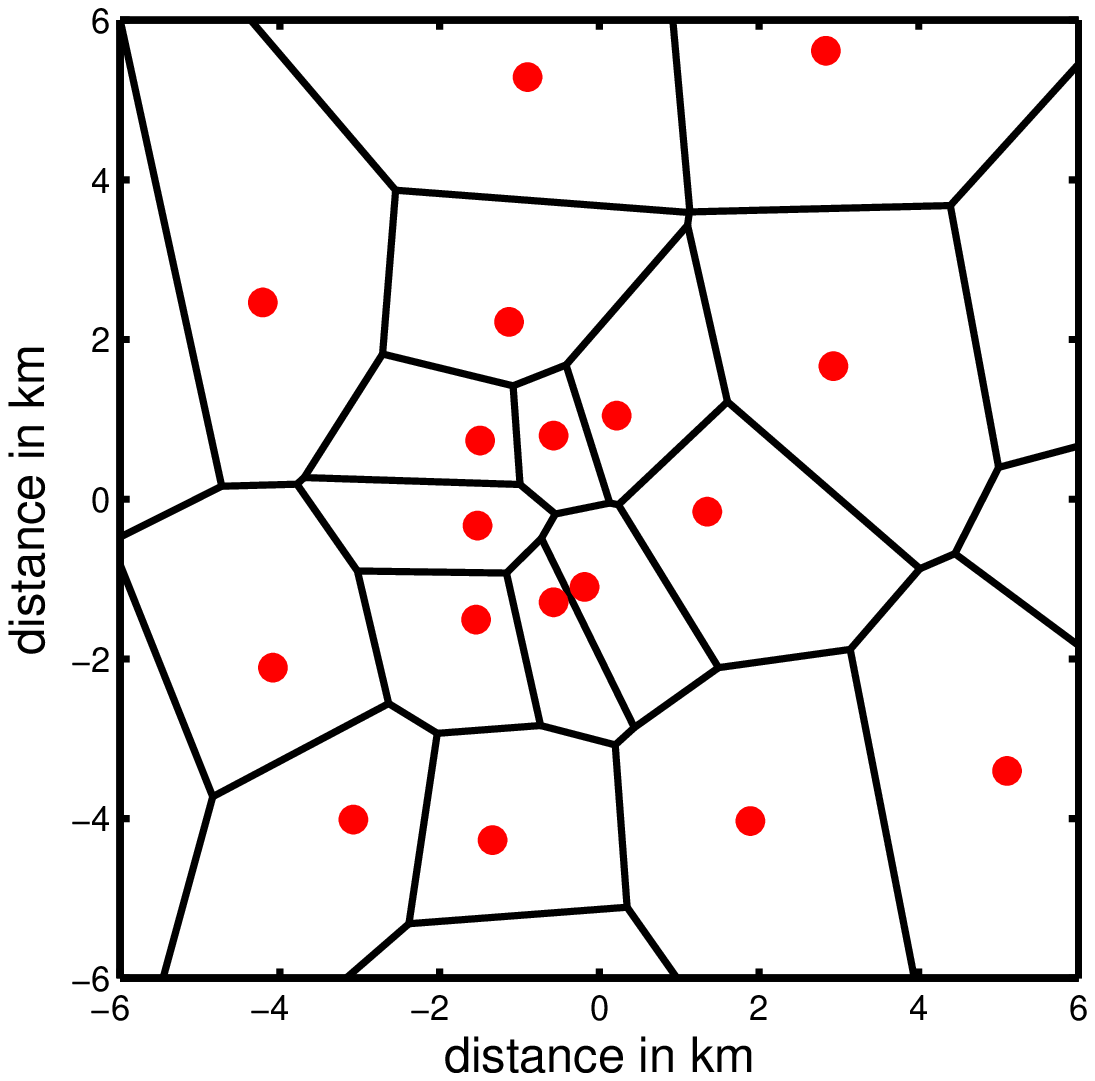}
&
\hspace{-0.5cm}
\includegraphics[width=4.55cm]{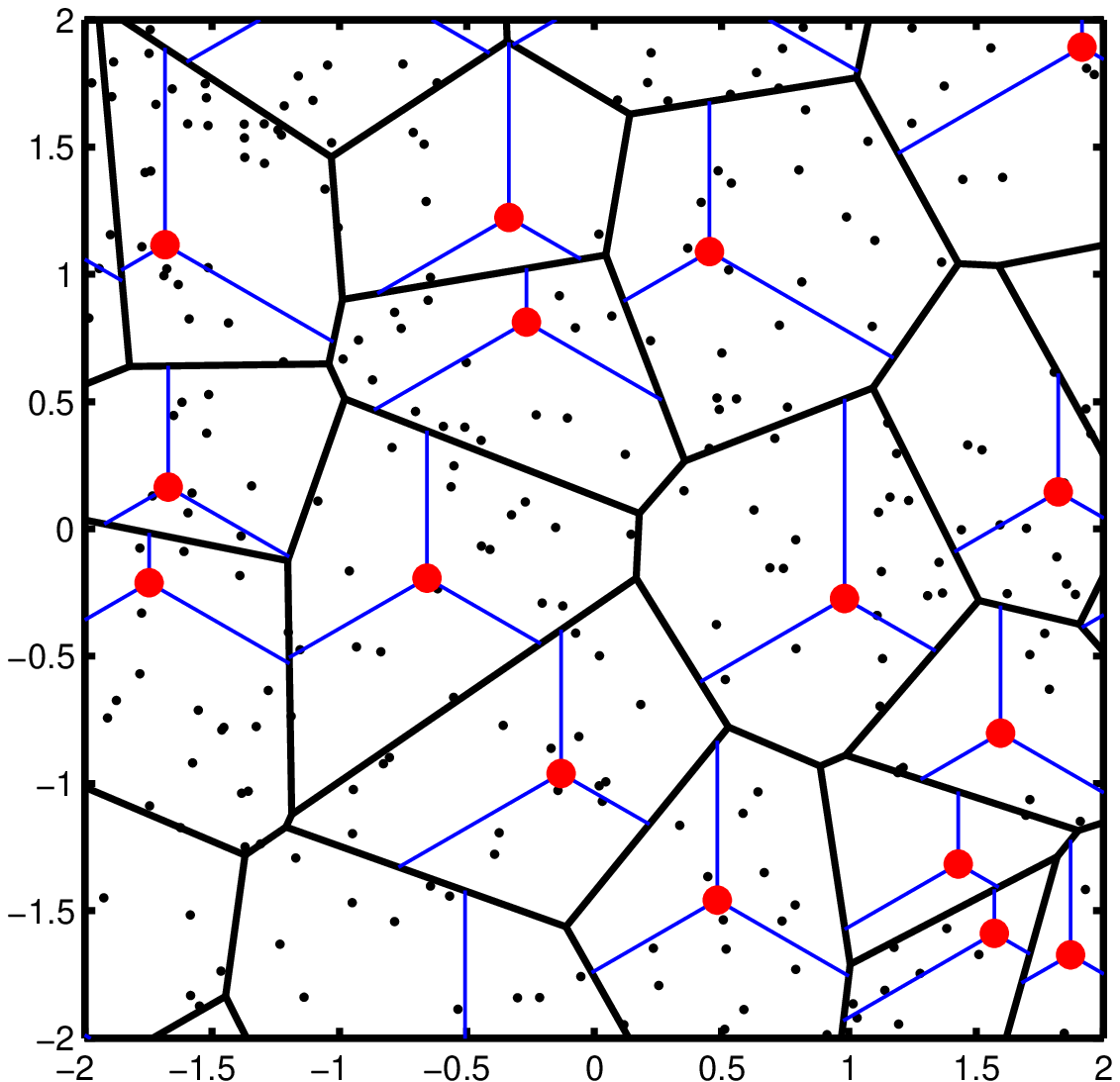}

\end{tabular}}

\vspace{-0.1cm}

\caption{Example network topologies. Base stations are represented by large
red circles, and cell boundaries are represented by solid black lines. Left
subfigure: Actual base station locations from a current cellular deployment.
Right subfigure: Simulated base-station locations using a base-station exclusion zone $r_\mathsf{bs} = 0.25$%
. In the right subfigure, the simulated positions of the mobiles are
represented by black dots, the sector boundaries are represented by light blue
lines, and the average cell load is $M/C=16$ mobiles. }
\label{Figure:Network}

\vspace{-0.5cm}

\end{figure}

Example network topologies are shown in Fig. \ref{Figure:Network}. The left
subfigure shows the locations of actual base stations in a small city with
hilly terrain. The base-station locations are given
by the large filled circles, and the Voronoi cells are indicated in the figure.
The minimum
base-station separation is observed to be about 0.43 km.
The right subfigure shows a portion of a
randomly generated network with average number of mobiles per cell $M/C=16$,
a base-station exclusion radius $r_{\mathsf{bs}}=0.25$, and a mobile
exclusion radius $r_{\mathsf{m}}=0.01$. The locations of the mobiles are
represented by small dots, and light lines indicate the angular
coverage of sector antennas.

Consider a reference receiver of a sector antenna that receives a desired
signal from a reference mobile within its cell and sector. Both intracell
and intercell interference are received from other mobiles within the
covered angle of the sector, but interference from mobiles in extraneous
sectors is negligible. Duplexing is assumed to prevent interference from
other sector antennas. The varying propagation delays from the interfering
mobiles cause their interference signals to be asynchronous with respect to
the desired signal.

In a DS-CDMA network of asynchronous quadriphase direct-sequence systems, a
multiple-access interference signal with power $I$ before despreading is
reduced after despreading to the power level $Ih(\tau_o)/G$, where G is
the processing gain or spreading factor, and $h(\tau_{o})$ is a function of
the chip waveform and the timing offset $\tau _{o}$ of the interference
spreading sequence relative to that of the desired or reference signal. If $%
\tau _{o}$ is assumed to have a uniform distribution over [0, $T_{c}],$ then
the expected value of $h(\tau _{o})$ is the chip factor $h$ . For
rectangular chip waveforms, $h=2/3$ \cite{tor}, \cite{tor3}. It is assumed
henceforth that $G/h(\tau _{o})$ is a constant equal to $G/h$ at each sector
receiver in the network.

Let $\mathcal{A}_{j}$ denote the set of mobiles \emph{covered} by sector
antenna $S_{j}$. A mobile $X_i \in \mathcal{A}_j$ will be \emph{associated}
with $S_j$ if the mobile's signal is received at $S_j$ with a higher average
power than at any other sector antenna in the network.
Let ${\mathcal{X}}_{j} \subset \mathcal{A}_j$ denote the set of mobiles
associated with sector antenna $S_{j}$.
Let $X_{r}\in \mathcal{X}_{j}$ denote a reference mobile that transmits a
desired signal to $S_{j}$. The power of $X_{r}$ received at $S_{j}$ is not
significantly affected by the spreading factor and depends on the fading and
path-loss models. The power of $X_{i},$ $i\neq r,$ received at $S_{j}$ is
nonzero only if $X_{i}\in \mathcal{A}_{j},$ is reduced by $G/h,$ and also
depends on the fading and path-loss models. We assume that path loss has a
power-law dependence on distance and is perturbed by shadowing. When
accounting for fading and path loss, the despread instantaneous power of $%
X_{i}$ received at $S_{j}$ is
\begin{equation}
\rho _{i,j}=
\begin{cases}
{P}_{r}g_{r,j}10^{\xi _{r,j}/10}f\left( ||S_{j}-X_{r}||\right) &
i=r \\
\left( \frac{h}{G}\right) {P}_{i}g_{i,j}10^{\xi _{i,j}/10}f\left(
||S_{j}-X_{i}||\right) & \hspace{-0cm} i:X_{i}\in \mathcal A_{j} \backslash X_r \\
0 & i:X_i \notin \mathcal A_j
\end{cases}
\label{eqn:power}
\end{equation}%
where $g_{i,j}$ is the power gain due to fading, $\xi _{i,j}$ is a \textit{%
shadowing factor}, ${P}_{i}$ is the power transmitted by $X_{i}$,
and $\mathcal A_j \backslash X_r$ is set $\mathcal A_j$ with element $X_r$ removed
(required since $X_r$ does not interfere with itself). The \{$%
g_{i,j}\}$ are independent with unit-mean, and $g_{i,j}=a_{i,j}^{2}$, where $%
a_{i,j}$ is Nakagami with parameter $m_{i,j}$. While the $\{g_{i,j}\}$ are
independent from each mobile to each base station, they are not necessarily
identically distributed, and each link can have a distinct Nakagami
parameter $m_{i,j}$. When the channel between $S_{j}$ and $X_{i}$
experiences Rayleigh fading, $m_{i,j}=1$ and $g_{i,j}$ is exponentially
distributed. In the presence of log-normal shadowing, the $\{\xi _{i,j}\}$
are i.i.d. zero-mean Gaussian random variables with variance $\sigma _{s}^{2}
$. In the absence of shadowing, $\xi _{i,j}=0$. \ The path-loss function is
expressed as the attenuation power law
\begin{equation}
f\left( d\right) =\left( \frac{d}{d_{0}}\right) ^{-\alpha },\text{ \ }d\geq
d_{0}  \label{eqn:pathloss}
\end{equation}%
where $\alpha \geq 2$ is the attenuation power-law exponent, and $d_{0}$ is
sufficiently large that the signals are in the far field. It is assumed that
$r_{\mathsf{bs}}$ and $r_{\mathsf{m}}$ exceed $d_{0}.$

It is assumed that the \{$g_{i,j}\}$ remain fixed for the duration of a time
interval, but vary independently from interval to interval (block fading).
With probability $p_{i}$, the $i^{th}$ interferer transmits in the same time
interval as the reference signal. The \emph{activity probability} $\{p_{i}\}$
can be used to model voice-activity factors or controlled silence. Although
the $\{p_{i}\}$ need not be the same, it is assumed that they are identical
in the subsequent examples.

Let $\mathsf{g}(i)$ denote a
function that returns the index of the sector antenna serving $X_{i}$ so
that $X_{i}\in {\mathcal{X}}_{j}$ if $\mathsf{g}\left( i\right) =j$.
Usually, the sector antenna $S_{\mathbf{g}\left( i\right) }$ that serves
mobile $X_{i}$ is selected to be the one with index
\begin{equation}
\mathsf{g}\left( i\right) =\underset{j}{\mathrm{argmax}}\,\left\{ 10^{\xi
_{i,j}/10}f\left( ||S_{j}-X_{i}||\right) ,\text{ }X_{i}\in \mathcal{A}%
_{j}\right\}
\end{equation}%
which is the sector antenna with minimum path loss from $X_{i}$ among those
that cover $X_{i}$. In the absence of shadowing, it will be the sector
antenna that is closest to $X_{i}$. In the presence of shadowing, a mobile
may actually be associated with a sector antenna that is more distant than
the closest one if the shadowing conditions are sufficiently better.

The instantaneous SINR at sector antenna $S_{j}$ when the desired signal is
from $X_{r}\in \mathcal{X}_{j}$ is
\begin{equation}
\gamma _{r,j}=\frac{\rho _{r,j}}{\displaystyle{\mathcal{N}}+
\sum_{i=1,i\neq r}^M  I_{i}\rho _{i,j}}  \label{eqn:SINR1}
\end{equation}%
where $\mathcal{N}$ is the noise power and $I_{i}$ is is a Bernoulli
variable with probability $P[I_{i}=1]=p_{i}$ and $P[I_{i}=0]=1-p_{i}$.
Substituting (\ref{eqn:power}) and (\ref{eqn:pathloss}) into (\ref{eqn:SINR1}%
) yields 
\begin{equation}
\gamma _{r,j}=\frac{g_{r,j}\Omega _{r,j}}{\displaystyle\Gamma ^{-1}
+\sum_{i=1,i\neq r}^MI_{i}g_{i,j}\Omega _{i,j}}
\end{equation}%
where $\Gamma =d_{0}^{\alpha }P_{r}/\mathcal{N}$ is the signal-to-noise
ratio (SNR) due to a mobile located at unit distance when fading and
shadowing are absent, and 
\begin{eqnarray}
\Omega _{i,j} &=&%
\begin{cases}
10^{\xi _{r,j}/10}||S_{j}-X_{r}||^{-\alpha } & i=r \\
\displaystyle\frac{hP_{i}}{GP_{r}}10^{\xi _{i,j}/10}||S_{j}-X_{i}||^{-\alpha
} & i:X_{i}\in \mathcal{A}_{j}\backslash X_{r} \\
0 & i:X_{i}\notin \mathcal{A}_{j}%
\end{cases}
\notag  \label{omega} \\
&&
\end{eqnarray}%
is the normalized mean despread power of $X_{i}$ received at $S_{j}$, where
the normalization is by $P_{r}$. The set of $\{\Omega _{i,j}\}$ for
reference receiver $S_{j}$ is represented by the vector $\boldsymbol{\Omega }%
_{j}=\{\Omega _{1,j},...,\Omega _{M,j}\}$.


\section{Outage Probability}

\label{Section:Outage}

Let $\beta _{r}$ denote the minimum instantaneous SINR required for reliable
reception of a signal from $X_{r}$ at its serving sector antenna $S_{j},j=%
\mathbf{g}(r)$. An \emph{outage} occurs when the SINR of a signal from $X_{r}
$ falls below $\beta _{r}$. The value of $\beta _{r}$ is a function of the
rate $R_{r}$ of the uplink, which is expressed in units of bits per channel
use (bpcu). The relationship $R_{r}=C(\beta _{r})$ depends on the modulation
and coding schemes used, and typically only a discrete set of $R_{r}$ can be
selected. While the exact dependence of $R_{r}$ on $\beta _{r}$ can be
determined empirically through tests or simulation, we make the simplifying
assumption when computing our numerical results that $C(\beta _{r})=\log
_{2}(1+\beta _{r})$ corresponding to the Shannon capacity for complex
discrete-time AWGN channels. This assumption is fairly accurate for modern
cellular systems such as Wideband CDMA (WCDMA) or LTE, which use turbo codes
with a large number of available rates.

Conditioning on $\boldsymbol{\Omega }_{j}$, the outage probability of a
desired signal from $X_{r}\in {\mathcal{X}}_{j}$ that arrives at $S_{j}$ is
\begin{equation}
\epsilon _{r}=P\left[ \gamma _{r,j}\leq \beta _{r}\big|\boldsymbol{\Omega }%
_{j}\right] .
\end{equation}%
Because it is conditioned on $\boldsymbol{\Omega }_{j}$, the outage
probability depends on the particular network realization, which has
dynamics over timescales that are much slower than the fading. In \cite{tor2}%
, it is proved that
\begin{equation}
\epsilon _{r}=1-e^{-\beta _{0}z}\sum_{s=0}^{m_{0}-1}{\left( \beta
_{0}z\right) }^{s}\sum_{t=0}^{s}\frac{z^{-t}H_{t}(\boldsymbol{\Psi })}{(s-t)!%
}
\end{equation}%
where $m_{0}=m_{r,j}$ is an integer, $\beta _{0}=\beta m_{0}/\Omega _{r,j}$,
\begin{align}
\Psi _{i}& =\left( \beta _{0}\frac{\Omega _{i,j}}{m_{i,j}}+1\right) ^{-1}%
\hspace{-0.5cm},\hspace{1cm}\mbox{for $i=\{1,...,M\}$, } \\
H_{t}(\boldsymbol{\Psi })& =\mathop{ \sum_{\ell_i \geq 0}}%
_{\sum_{i=0}^{M}\ell _{i}=t}\prod_{i=1,i\neq r}^{M}G_{\ell _{i}}(\Psi _{i}),
\end{align}%
\begin{equation}
G_{\ell }(\Psi _{i})=%
\begin{cases}
1-p_{i}(1-\Psi _{i}^{m_{i,j}}) & \mbox{for $\ell=0$} \\
\frac{p_{i}\Gamma (\ell +m_{i,ji})}{\ell !\Gamma (m_{i,j})}\left( \frac{%
\Omega _{i,j}}{m_{i,j}}\right) ^{\ell }\Psi _{i}^{m_{i,j}+\ell } &
\mbox{for
$\ell>0$.}%
\end{cases}%
.
\end{equation}

\section{Network Policies\label{Section:Policies}}

\subsection{Power Control}

\label{pc} A standard power-allocation policy for DS-CDMA networks is to
select the transmit power $\{P_{i}\}$ for all mobiles in the set $\mathcal{X}%
_{j}$ such that, after compensation for shadowing and power-law attenuation,
each mobile's transmission is received at sector antenna $S_{j}$ with the
same power $P_{0}$.
For such a power-control policy, each mobile in $\mathcal{X}_{j}$ will
transmit with a power $P_{i}$ that satisfies
\begin{equation}
{P}_{i}10^{\xi_{i,j}/10}f\left( ||S_{j}-X_{i}||\right) =P_{0},\text{ \ }%
X_{i}\in \mathcal{X}_{j}.  \label{Eqn:Intracell}
\end{equation}
Since the reference mobile $X_r \in \mathcal{X}_{j}$, its transmit power is
determined by (\ref{Eqn:Intracell}).

For a reference mobile $X_{r}$, the interference at sector antenna $S_{j}$
is from the mobiles in the set $\mathcal{A}_{j}\backslash X_{r}$. This set
can be partitioned into two subsets. The first subset $\mathcal{X}%
_{j}\backslash X_{r}$ comprises the \textit{intracell interferers}, which
are the other mobiles in the same cell and sector as the reference mobile.
The second subset $\mathcal{A}_{j}\backslash \mathcal{X}_{j}$ comprises the
\textit{intercell interferers}, which are the mobiles covered by sector
antenna $S_{j}$ but associated with a cell sector other than $\mathcal{X}%
_{j} $.

Considering intracell interference, all of the mobiles within the sector
transmit with power given by (\ref{Eqn:Intracell}). Substituting (\ref%
{Eqn:Intracell}) and (\ref{eqn:pathloss}) into $\left( \ref{omega}\right) $,
we obtain the normalized received power of the intracell interferers
\begin{equation}
\Omega _{i,j}=\frac{h}{G}10^{\xi _{r,j}/10}||S_{j}-X_{r}||^{-\alpha },\text{
\ }X_{i} \in \mathcal{X}_{j} \backslash X_r.
\end{equation}%
Although the number of mobiles in the reference cell sector must be
known to compute the outage probability, the locations of these mobiles in
the cell are irrelevant to the computation of the $\Omega _{i,j}$ of the
intracell interferers.

Considering intercell interference, the set $\mathcal{A}_{j}\backslash
\mathcal{X}_{j}$ can be further partitioned into sets $\mathcal{A}_{j}\cap
\mathcal{X}_{k}$, $k\neq j$, containing the mobiles covered by sector
antenna $S_{j}$ but associated with some other sector antenna $S_{k}$. For
those mobiles in $\mathcal{A}_{j}\cap \mathcal{X}_{k}$, power control
implies that
\begin{equation}
{P}_{i}10^{\xi _{i,k}/10}f\left( ||S_{k}-X_{i}||\right) =P_{0},\text{ \ }%
X_{i}\in \mathcal{X}_{k}\cap A_{j},\text{ \ }k\neq j.  \label{Eqn:Inter}
\end{equation}%
Substituting (\ref{Eqn:Inter}), $\left( \ref{Eqn:Intracell}\right) $, and (%
\ref{eqn:pathloss}) into $\left( \ref{omega}\right) $ yields%
\begin{eqnarray}
\Omega _{i,j} &=&\frac{h}{G}10^{\xi _{i,j}^{\prime }/10}\left( \frac{%
||S_{j}-X_{i}||||S_{j}-X_{r}||}{||S_{k}-X_{i}||}\right) ^{-\alpha }\text{ \ }
\notag \\
\xi _{i,j}^{\prime } &=&\xi _{i,j}+\xi _{r,j}-\xi _{i,k},\text{ \ }X_{i}\in
\mathcal{X}_{k}\cap A_{j},\text{ \ }k\neq j\text{ \ }
\end{eqnarray}%
for $\mathcal{A}_{j}\backslash \mathcal{X}_{j}$, which gives the mean
intercell interference power at the reference sector antenna due to
interference from mobile $i$ of sector $k=\mathsf{g}(i)$. Assuming that
imperfect power control is caused by numerous independent multiplicative
effects, it may be approximately modeled by increasing the standard
deviations of $\xi _{i,j},$ $\xi _{r,j},$ and $\xi _{i,k}.$

\subsection{Rate Control}

\label{Section:RateControl}

In addition to controlling the transmitted power, the rate $R_{i}$ of each
uplink needs to be selected. Due to the irregular network geometry, which
results in cell sectors of variable areas and numbers of intracell mobiles,
the amount of interference received by a sector antenna can vary
dramatically from one sector to another. With a fixed rate for each sector,
the result is a highly variable outage probability. An alternative to using
a fixed rate for the entire network is to adapt the rate of each uplink to
satisfy an outage constraint or maximize the throughput of each uplink. Many
networks adapt the modulation and coding to increase the throughput in
accordance with the channel conditions. For example, in an 802.11g network,
more than ten combinations of modulation and coding are available, including
QPSK, 16-QAM, and 64-QAM.

To illustrate the influence of rate on performance, consider the following
example. The network has $C=50$ base stations and $M=400$ mobile stations
placed in a circular network of radius $r_{\mathsf{net}}=2$. The
base-station exclusion zones have radius $r_{\mathsf{bs}}=0.25$, while the
mobile exclusion zones have radius $r_{\mathsf{m}}=0.01$. The spreading factor
is $G=16,$ and the chip factor is $h=2/3$. Since $M/C=G/2$, the network
is characterized as being \emph{half loaded}. The SNR is $\Gamma =10$ dB,
the activity factor is $p_{i}=1$, the path-loss exponent is $\alpha =3$, and
shadowing is assumed with $\sigma _{s}=8$ dB. A \emph{distance-dependent
fading} model is assumed, where $m_{i,j}$ is set according to:
\begin{equation}
m_{i,j}=%
\begin{cases}
3 & \mbox{ if }\;||S_{j}-X_{i}||\leq r_{\mathsf{bs}}/2 \\
2 & \mbox{ if }\;r_{\mathsf{bs}}/2<||S_{j}-X_{i}|\leq r_{\mathsf{bs}} \\
1 & \mbox{ if }\;||S_{j}-X_{i}|>r_{\mathsf{bs}}%
\end{cases}%
.
\label{eqn:ddfading}
\end{equation}%
The distance-dependent-fading model characterizes the situation where a
mobile close to the base station is in the line-of-sight, while mobiles
farther away tend to be non-LOS.

\begin{figure}[t]
\centering
\includegraphics[width=8.5cm]{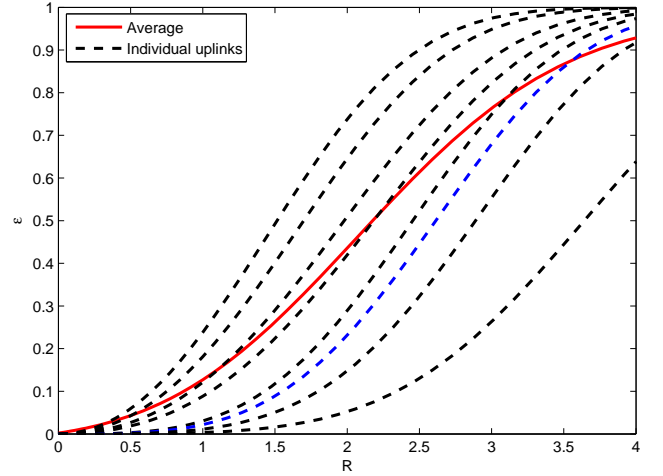} 
\vspace{-0.25cm}
\caption{ Outage probability of eight randomly selected uplinks (dashed
lines) along with the average outage probability for the entire network
(solid line). The results are for a half-loaded network ($M/C=G/2$), with
distance-dependent fading and shadowing ($\protect\sigma_s=8$ dB) and are
shown as a function of the rate $R$. }
\label{Figure:OutagePC}
\vspace{-0.5cm}
\end{figure}

Fig. \ref{Figure:OutagePC} shows the outage probability as a function of
rate. Assuming the use of a capacity-approaching code, two-dimensional
signaling over an AWGN channel, and Gaussian interference, the SINR
threshold corresponding to rate $R$ is $\beta =2^{R}-1$. The dashed lines in
Fig. \ref{Figure:OutagePC} were generated by selecting eight random uplinks
and computing the outage probability for each using this threshold. Despite
the use of power control, there is considerable variability in the outage
probability. The outage probabilities \{$\epsilon _{i}\}$ were computed for
all $M$ uplinks in the system, and the average outage probability,
\begin{equation}
\mathbb{E}[\epsilon ]=\frac{1}{M}\sum_{i=1}^{M}\epsilon _{i}
\end{equation}%
is displayed as a solid line in the figure.

\begin{figure}[t]
\centering
\includegraphics[width=8.5cm]{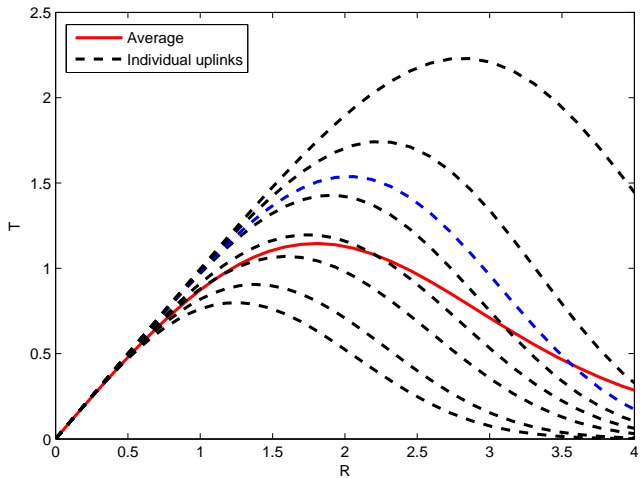}
\vspace{-0.25cm}
\caption{ Throughput of eight randomly selected uplinks (dashed lines) along
with the average throughput for the entire network (solid line). System
parameters are the same used to generate Fig. \protect\ref{Figure:OutagePC}.
}
\label{Figure:TCPC}
\vspace{-0.25cm}
\end{figure}

Fig. \ref{Figure:TCPC} shows the \emph{throughput} as a function of rate,
where the throughput of the $i^{th}$ uplink is found as
\begin{equation}
T_{i}=R_{i}(1-\epsilon _{i})
\end{equation}%
and represents the rate of successful transmissions. The parameters are the
same as those used to produce Fig. \ref{Figure:OutagePC}, and again the SINR
threshold corresponding to rate $R$ is $\beta =2^{R}-1$. The plot shows the
throughput for the same eight uplinks whose outage was shown in Fig. \ref%
{Figure:OutagePC}, as well as the average throughput
\begin{equation}
\mathbb{E}[T]=\frac{1}{M}\sum_{i=1}^{M}R_{i}(1-\epsilon _{i}).
\end{equation}

\subsubsection{Fixed-Rate Policies}

Consider the case that all uplinks in the system must use the same rate;
i.e., $R_{i}=R$, for all uplinks. On the one hand, the rate could be
selected to maximize the average throughput. With respect to the example
shown in Fig. \ref{Figure:TCPC}, this corresponds to selecting the $R$ that
maximizes the solid curve, which occurs at $R=1.81$. However, at the rate
that maximizes throughput, the corresponding outage probability could be
unacceptably high. When $R=1.81$ in the example, the corresponding average
outage probability is $\mathbb{E}[\epsilon ]=0.37$, which is too high for
many applications. As an alternative to maximizing throughput, the rate $R$
could be selected to satisfy an outage constraint $\zeta $ so that $\mathbb{E%
}[\epsilon ]\leq \zeta $. For instance, setting $R=0.84$ in the example
satisfies an average outage constraint $\zeta =0.1$ with equality. To
distinguish between the two fixed-rate policies, we call the first policy
\emph{maximal-throughput fixed rate} (MTFR) and the second policy \emph{%
outage-constrained fixed rate} (OCFR).

\subsubsection{Variable-Rate Policies}

If $R$ is selected to satisfy an average outage constraint, the outage
probability of the individual uplinks will vary around this average.
Furthermore, selecting $R$ to maximize the average throughput does not
generally maximize the throughput of the individual uplinks. These issues
can be alleviated by selecting the rates $R_{i}$ independently for the
different uplinks. The rates could be selected to maximize the throughput of
each uplink; i.e., $R_{i}=\arg \max T_{i}$ for each uplink, where the
maximization is over all possible rates. We call this the \emph{%
maximal-throughput variable-rate} (MTVR) policy. Alternatively, the
selection could be made to require all uplinks to satisfy the outage
constraint $\zeta $; i.e., $\epsilon _{i}\leq \zeta $ for all $i$. We call
this the \emph{outage-constrained variable-rate} (OCVR) policy.

\section{Performance Analysis}

\label{Section:Performance}

\subsection{Performance Metrics}

While the outage probability, throughput, and rate characterize the
performance of a single uplink, they do not quantify the total data flow in
the network because they do not account for the number of uplink users that
are served. By taking into account the number of mobiles per unit area, the
total data flow in a given area can be characterized by the \emph{%
transmission capacity}, defined as \cite{web}
\begin{equation}
\tau =\lambda \mathbb{E}[T]=\lambda \mathbb{E}\left[ \left( 1-\epsilon
\right) R\right]  \label{eqn:tc}
\end{equation}%
where
$\lambda =M/\pi r_{\mathsf{net}}^{2}$ is the density of transmissions in the
network, and the units are bits per channel use per unit area. Transmission
capacity can be interpreted as the spatial intensity of transmissions; i.e.,
the rate of successful data transmission per unit area. 

Performance metrics can be obtained using
a Monte Carlo approach as follows. Draw a realization of the network by
placing $C$ base stations and $M$ mobiles within the disk of radius $r_{%
\mathsf{net}}$ according to the uniform clustering model with minimum
base-station separation $r_{\mathsf{bs}}$ and minimum mobile separation $r_{%
\mathsf{m}}$. Compute the path loss from each base station to each mobile,
applying randomly generated shadowing factors. Determine the set of mobiles
associated with each cell sector. Assuming that the number of mobiles served
in a cell sector cannot exceed $G$, which is the number of orthogonal
sequences available for the downlink, then set the rate of the last $M_{j}-G$
mobiles in the cell sector (if there are that many) to zero (since they will
be denied access). At each sector antenna, apply the power-control policy to
determine the power it receives from each mobile that it serves. For each
cell sector, compute the rates, outage probabilities, and throughputs
according to the MTFR, OCFR, MTVR, or OCVR network policy. Then compute the
transmission capacity.

\subsection{Simulation Parameters}

In the following subsections, the Monte Carlo method described in the
previous subsection is used to characterize the uplink performance. In all
cases considered, the network has $C=50$ base stations placed in a circular
network of radius $r_\mathsf{net}=2$. Except for Subsection \ref{Section:RBS}%
, which studies the influence of $r_\mathsf{bs}$, the base-station exclusion
zones are set to have radius $r_\mathsf{bs}=0.25$. A variable number $M$ of
mobiles are placed within the network using exclusion zones of radius $r_%
\mathsf{m}=0.01$. The SNR is $\Gamma =10$ dB, and the activity factor is $%
p_i=1$. Unless otherwise stated, the path-loss exponent is $\alpha = 3$.
The fading is the distance-dependent fading specified by (\ref{eqn:ddfading}).
Both unshadowed and shadowed ($%
\sigma_s = 8$ dB) environments are considered. The chip factor is $h=2/3$,
and except for Subsection \ref{Section:PG}, which studies the influence of $%
G $, the spreading factor is $G = 16$.

\subsection{Policy Comparison}

\begin{figure}[t]
\centering
\includegraphics[width=8.5cm]{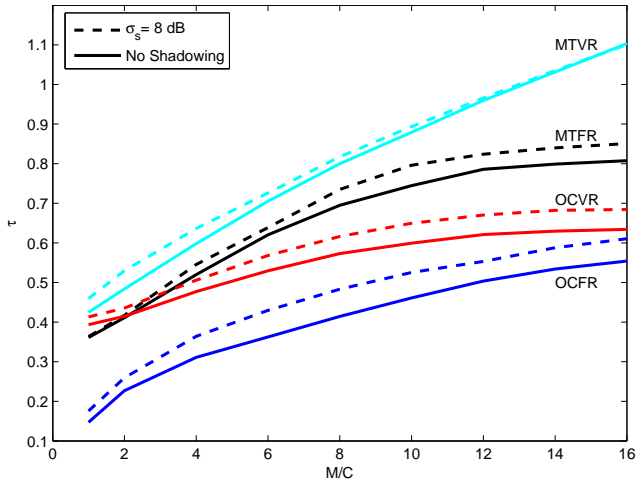} 
\vspace{-0.25cm}
\caption{ Transmission capacity for the four network policies as function of
the load $M/C$ for distance-dependent fading and both shadowed ($\protect%
\sigma _{s}=$ 8 dB) and unshadowed cases, assuming that the rates are
optimized. }
\vspace{-0.4cm}
\label{Figure:TCComp}
\end{figure}

Fig. \ref{Figure:TCComp} shows the average transmission capacity of the four
network policies in distance-dependent fading, both with and without
shadowing, as a function of the load $M/C$. For the OCFR policy, the average
was obtained by optimizing over a rate that is common to all sectors. For
the OCVR policy, the uplink rate $R_{i}$ of each uplink was maximized
subject to an outage constraint. While the transmission capacities of the
MTFR and MTVR policies are potentially superior to those of the OCFR and
OCVR policies, this advantage comes at the cost of a variable and high value
of $\epsilon $, which is generally too large for most applications. The
bottom pair of curves in Fig. \ref{Figure:TCComp} indicate that the OCVR\
policy has a higher average transmission capacity than the OCFR policy.

\subsection{ Spreading Factor }

\label{Section:PG}

\begin{figure}[t]
\centering
\includegraphics[width=8.5cm]{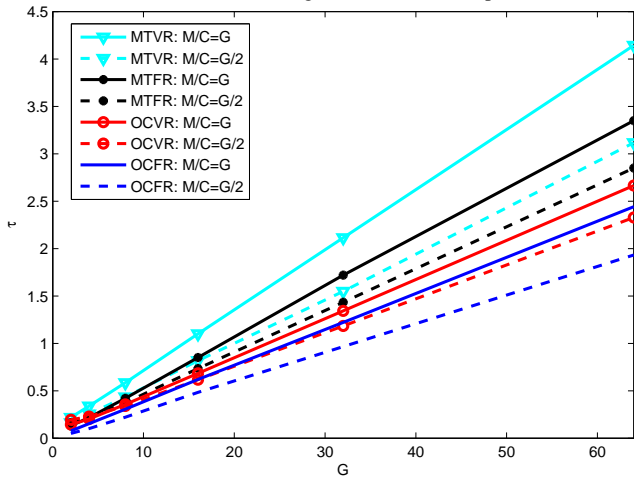}
\vspace{-0.25cm}
\caption{ Transmission capacity as function of spreading factor $G$ for two
values of system load, distance-dependent fading, and shadowing with $%
\protect\sigma _{s}=8$ dB.
}
\vspace{-0.75cm}
\label{Figure:PG}
\end{figure}

Fig. \ref{Figure:PG} shows the transmission capacity as a function of the
spreading factor $G$ (with $h=2/3$) for the shadowed
distance-dependent-fading channel. Two loads are shown for each of the four
policies. An increase in $G$ is beneficial for all policies, but the MTVR
policy benefits the most.

\subsection{ Base-Station Exclusion Zone}

\label{Section:RBS}

\begin{figure}[t]
\centering
\includegraphics[width=8.5cm]{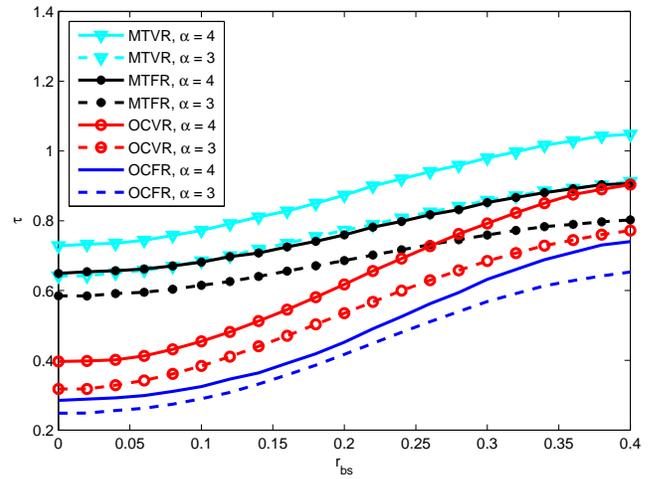}
\vspace{-0.25cm}
\caption{ Transmission capacity as a function of the base-station exclusion
zone $r_{\mathsf{bs}}$ for four policies and two values of path-loss
exponent $\protect\alpha $.}
\vspace{-0.4cm}
\label{Figure:TCRbs}
\end{figure}

Fig. \ref{Figure:TCRbs} shows the transmission capacity for each of the four
policies as a function of the base-station exclusion zone $r_{\mathsf{bs}}$
for $M/C=G/2$ and two values of path-loss exponent $\alpha $. The
distance-dependent fading model is used, and shadowing is applied with $%
\sigma _{s}=$ 8 dB. The two policies that constrain the outage probability
are more sensitive to the value of $r_{\mathsf{bs}}$ than the two policies
that maximize throughput.

\section{ Conclusion}

A new analysis of DS-CDMA uplinks has been presented. This analysis is much
more detailed and accurate than existing ones and facilitates the resolution
of network design issues. In particular, it has been shown that once power
control is established, the rate can be allocated according to a fixed-rate
or variable-rate policy with the objective of either maximizing throughput
or meeting an outage constraint. An advantage of variable-rate power control
is that it allows an outage constraint to be enforced on every uplink, which
is impossible when a fixed rate is used throughout the network. Another
advantage is an increased transmission capacity.

\balance


\end{document}